\documentclass[12pt,a4paper]{article}  

\usepackage{amssymb}
\usepackage{amsmath}
\usepackage{latexsym}
\usepackage{bbold}
\usepackage{mathbbol}
\usepackage{hyperref}

\newcommand{\be}{\begin{equation}}
\newcommand{\ee}{\end{equation}}
\newcommand{\bea}{\begin{eqnarray}}
\newcommand{\eea}{\end{eqnarray}}

\newcommand{\de}{\partial}
\newcommand{\ket}[1]{\lvert #1 \rangle} 
\newcommand{\Dslash}{D \! \! \!  \!  /} 

\newcommand{\media}[1]{\langle #1 \rangle} 
\newcommand{\Det}{\mathrm{Det}}

\begin{document}

\begin{titlepage}

\begin{center}

{\Large\bf Particles with non abelian charges}
\vskip 1.2cm

Fiorenzo Bastianelli$^{\,a}$, Roberto Bonezzi$^{\,a}$, Olindo Corradini$^{\,b}$, Emanuele Latini$^{\,c}$ 

\vskip 1cm
$^a${\em Dipartimento di Fisica e Astronomia, Universit{\`a} di Bologna and\\
INFN, Sezione di Bologna, via Irnerio 46, I-40126 Bologna, Italy}
\vskip .3cm
$^b${\em Centro de Estudios en F\'isica y Matem\'aticas Basicas y Aplicadas,\\
Universidad Aut\'onoma de Chiapas, Ciudad Universitaria, Tuxtla Guti\'errez 29050, Mexico}
\vskip .3cm
$^c${\em Institut f{\"u}r Mathematik, Universit{\"a}t Z{\"u}rich-Irchel, \\ Winterthurerstrasse 190, CH-8057 Z{\"u}rich, Switzerland}

\end{center}
\vskip .8cm

\abstract{Efficient methods for describing non abelian charges in worldline approaches to QFT  are useful  to simplify 
calculations and address structural properties, as for example color/kinematics relations.
Here we analyze in detail a method  for treating arbitrary non abelian charges. We use Grassmann variables to take 
into account color degrees of freedom, which however are known to produce reducible representations of the color group. 
Then we couple them to a U(1) gauge field defined on the worldline, together with a Chern-Simons term,
to achieve  projection on an irreducible representation. Upon gauge fixing there remains a modulus, an angle parametrizing 
the U(1) Wilson loop, whose dependence is taken into account exactly in the propagator of the Grassmann variables.  
We test the method in simple examples, the scalar and spin 1/2 contribution to the gluon self energy, 
and suggest that it might simplify the analysis of more involved amplitudes.}
\vskip 2cm

\begin{center}
{\em Dedicated to the Memory of V\'ictor Manuel Villanueva Sandoval}
\end{center}

\end{titlepage}
\newpage

\section{Introduction and outlook}
The description of non abelian charges in worldline approaches has a long history. 
Yet, it seems useful to improve on  existing methods to gain in efficiency and be able to address in the worldline context 
properties like the color/kinematics relations,  found in the study of perturbative gauge and gravitational amplitudes  \cite{Bern:2008qj}.

The classical limit of nonabelian charges was originally studied in \cite{Wong:1970fu}, followed by the proposal of 
using Grassmann variables to treat them in first quantization \cite{Balachandran:1976ya, Barducci:1976xq}.
The usefulness of Grassmann variables is that upon quantization they produce a finite dimensional Hilbert space
that can be employed to describe both spin and color degrees of freedom. 
For the spin 1/2 particle the description of spin in terms of Grassmann  variables
\cite{Berezin:1976eg, Brink:1976sz}
allows to simplify in the electromagnetic coupling the spin factor 
introduced with a path ordering prescription by Feynman \cite{Feynman:1951gn},
so that the worldline description can combine into one expression
Feynman diagrams with different orderings of the external photons
along a line or loop, just as in the scalar case. This produces considerable 
simplifications for the organization and calculation of the related amplitudes.
However, the color Hilbert space arising from the use of Grassmann variables is reducible.
For that reason path ordered exponentials (i.e. Wilson lines) to take into account non abelian charges 
were often used in the past, as for example in the worldline calculations of 
\cite{Strassler:1992zr, Bastianelli:1992ct, Reuter:1996zm, Sato:1998sf, Schubert:2001he, 
Dai:2008bh, Ahmadiniaz:2012xp, Ahmadiniaz:2012ie}.
The quantization of Grassmann variables creates automatically the path ordering, but one must devise ways to select the irreducible 
representation needed in specific applications.
One way of projecting onto irreducible representations was proposed in  \cite{D'Hoker:1995bj} by considering a sum over a discrete set 
of angles, that implements the required projection. 
This method was used for example in \cite{JalilianMarian:1999xt} and \cite{Khorsand:2001dx}.

Here we suggest a way of obtaining the projection by coupling the Grassmann variables to a U(1) worldline 
gauge field with an additional Chern-Simon term, whose discretized coupling is tuned to obtain the required 
projection. This projection mechanism  appeared in the worldline description of differential forms
\cite{Howe:1989vn, Bastianelli:2005vk, Bastianelli:2011pe}, defined by constrained models with worldline gauge fields 
making up an extended supergravity multiplet with a U(1) gauge field. Upon quantization the latter is seen to produce
the projection needed to select the degree of the differential form. This mechanism was then used in other worldline 
applications \cite{Dai:2008bh, Bastianelli:2009eh, Bastianelli:2012nh,  Bastianelli:2013tsa}, where a suitable projection was needed.
After gauge fixing, there remains an integration over a modulus associated
to the U(1) gauge field,  an angle $\phi$, that implements the projection in the amplitudes.
In \cite{D'Hoker:1995bj} it was indeed suggested that the discrete sum used there to achieve projection 
could be turned into a continuous integration.
In our model we see how that suggestion is explicitly realized.
 To proceed further, we find it useful to encode the coupling 
of the Grassmann variables to the  modulus $\phi$ by using twisted boundary conditions.
One-loop amplitudes are then obtained as usual by computing worldline correlators of vertex operators
suitably integrated over the moduli $T$ and $\phi$, the proper time and the U(1) modulus.
Here we apply the mechanism to obtain projection on a given representation of the color group
for particles of spin 0 and 1/2, and as a test calculation compute their contribution to the gluon self energy. 
Gluons can also be described in first quantization, as in \cite{Strassler:1992zr, Schubert:2001he}
or with the string inspired method of  \cite{Dai:2008bh}, and one could again employ in such a case the present
description of non abelian charges.
It would be interesting to analyze how this description of color degrees of freedom performs in calculating
higher point one-loop amplitudes, and perhaps study the origin of the color/kinematics relations in this context.

A similar set up may be constructed using bosonic variables instead of fermionic ones.
The corresponding Hilbert space  that gives rise to the color degrees of freedom is infinite dimensional, 
containing all possible symmetric tensor products of the fundamental representation, but the coupling 
to the U(1) gauge field with a properly chosen Chern-Simons coupling can again
select the finite subspace corresponding to the fundamental representation.

\section{The colored scalar particle}
The euclidean action of a scalar particle coupled to the photon is given by
\be
S[x,e;A] = \int_0^1 d\tau \Big(\frac12 e^{-1} \dot x^2  + \frac12 e m^2 - i q A_\mu(x) \dot x^\mu\Big)
\ee
where $x^\mu$ are the coordinates of the particle, $e$ is the einbein, $q$ is the charge of the particle, 
and $A_\mu(x)$ is the background abelian gauge field. 
When inserted in the path integral
\be
\int  \  \frac{{D} x  {D} e}{\rm Vol(Gauge)}\ {e}^{- S[x,e;A]}  
\ee
one finds that the coupling gives rise to the Wilson line 
$
e^{\,i q \int A_\mu d x^\mu} 
$.  
A way to generalize this to non abelian fields $A_\mu (x) = A_\mu^a  (x) T^a$, with  $T^a$ hermitian generators of a  simple 
group, is to use the path ordering prescription to guarantee gauge invariance 
\be 
P\, {e}^{\, i g \int A_\mu d x^\mu}\;.
\ee
The generators are taken in an arbitrary representation of the gauge group and describe the non abelian charge assigned to the particle, 
while $g$ denotes the coupling constant. 

This procedure is correct and well-known,
but it is also useful to introduce Grassmann variables to create the Hilbert space associated to the  color degrees of freedom
and get rid of the path ordering prescription. The latter is generated by path integration over the new variables.
The coupling part of the scalar particle action takes then the form
\be
\Delta S_{_{NA}} = 
\int_0^1 d \tau  \bigg ( \bar c^\alpha \dot c_\alpha -i g A^a _\mu(x)\dot x^\mu  \bar c^\alpha (T^a)_\alpha{}^\beta c_\beta  \bigg ) 
\label{1.5}
\ee 
which depend on the Grassmann variables  $c_\alpha$ and their complex conjugates $\bar c^\alpha$.
The gauge group generators $(T^a)_\alpha{}^\beta$ can be chosen in any desired representation, but for definiteness we 
select SU($N$) as gauge group and choose the fundamental representation which has dimension $N$.
The Grassmann variables $c_\alpha$ has thus an index in
the fundamental representation, and   $\bar c^\alpha$ in the complex conjugate one.

The quantization of the Grassmann variables gives rise to fermionic creation and annihilation  operators
$\hat c_\alpha$ and $\hat c^{\dagger \alpha}$, satisfying the anticommutation relations
\be
\{ \hat c_\alpha, \hat c^{\dagger \beta}\} =\delta_\alpha^\beta\;, \quad  
\{ \hat c_\alpha, \hat c_\beta \} =0\;, \quad  
\{\hat c^{\dagger \alpha}, \hat c^{\dagger \beta}\} =0 \;.
\ee
They are naturally represented by $\hat c^{\dagger \alpha} \sim \bar c^\alpha$ and 
 $\hat c_\alpha \sim \frac{\partial }{\partial \bar c^\alpha}$
 when acting on wave functions of the form $\phi(x,\bar c)$. The latter has a finite Taylor expansion on the Grassmann
 numbers  $\bar c^\alpha$ of the form
 \be
 \phi(x,\bar c)= \phi(x) + \phi_\alpha(x)\bar c^\alpha +\frac12 \phi_{\alpha\beta}(x)\bar c^\alpha \bar c^\beta +\cdots
 +\frac{1}{N!} \phi_{\alpha_1 ...\alpha_N}(x)\bar c^{\alpha_1} ...\bar c^{\alpha_N}  
 \label{wavefun}
 \ee 
and contain  wave functions transforming in all possible antisymmetric tensor products of the fundamental representation.
Thus we see that the fermionic creation and annihilation  operators create a finite dimensional Hilbert space
for the color degrees of freedom, which is however reducible.

To select the fundamental representation one must project to the sector with occupation number one, 
so to isolate the wave function  $\phi_\alpha(x)$ with an index in the fundamental representation.
This can be achieved by coupling the variables $c_\alpha$ and $\bar c^\alpha$  to a U(1) gauge field $a(\tau)$ living on the worldline, 
with in addition a Chern-Simons term with quantized coupling $s =n- \frac{N}{2} $, where $n$  is the occupation number and
$N$ is the dimension of the fundamental representation.
The gauge field $a$ acts as a Lagrange multiplier that imposes a constraint on physical states, and setting $n=1$ selects precisely
the sector with occupation number one as possible physical states,
corresponding to wave functions in the fundamental representation of the color group.

Let us analyze this mechanism in detail. The free kinetic term of the Grassmann variables that appears  in (\ref{1.5}) is  modified 
by the coupling to the worldline U(1) gauge field $a$ to
\be
S_{c\bar c }  =
\int_0^1 d \tau \Big (\bar c^\alpha (\partial_\tau + i  a ) c_\alpha  - i s a \Big) \ee
where the last piece is the Chern-Simons term.
The equations of motion of $a$ produce the constraint 
$C\equiv \bar c^\alpha c_\alpha - s = 0$. Upon quantization the function $C$ 
becomes the operator
\be
\hat C \equiv \frac12 ( \hat c^{\dagger \alpha} \hat c_\alpha  - \hat c_\alpha \hat c^{\dagger \alpha})  - s  
\ \sim \
  \bar c^\alpha \frac{\partial }{\partial \bar c^\alpha} -1
\ee
where we have resolved an ordering ambiguity symmetrizing   in a graded way the term with $c$ and $\bar c$,
and used the quantized value of the Chern-Simons coupling $s= 1-\frac{N}{2}$.
Clearly, when acting on the generic wave function (\ref{wavefun}) the constraint
$\hat C \phi(x, \bar c) =0$ selects the wave function $\phi_\alpha(x)$.
The conclusion is that one can compute quantum properties of a scalar particle
coupled to non abelian gauge fields in an arbitrary representation by using the above ingredients. 

In particular,  the one-loop effective action induced by a scalar particle coupled to a non abelian gauge field is computed  by
path integrating on the circle $S^1$
\be
\Gamma[A] = \int_{S^1}  \ 
\frac{{D} x  {D} \bar c  {D} c {D}  e {D}  a}{ {\rm Vol(Gauge)}}\
{e}^{- S}  
\ee
where
\be
S=
 \int_0^1 d\tau \bigg(\frac12 e^{-1}
 \dot x^2 + \frac12 e m^2  
 + \bar c^\alpha \big (\partial_\tau + i a\big ) c_\alpha  - i s a
-ig A^a _\mu(x)\dot x^\mu  \bar c^\alpha (T^a)_\alpha{}^\beta c_\beta \bigg)\;.
\ee
The particle action has two local invariances, the standard reparametrization invariance and the new U(1) local symmetry.
One may gauge fix $e(\tau)=2T$ and $a(\tau)=\phi$, where $T$ is the standard Fock-Schwinger proper time 
and $\phi$ an angle describing U(1) gauge invariant configurations on the circle
($ z=e^{i \int_0^1d\tau a} 
= e^{i\phi}$ is the Wilson loop).
They make up moduli that must be integrated over in the path integral. Taking care of the related Faddeev-Popov determinants,
one finds a final formula for the induced QFT effective action for the non abelian field $A^a_\mu$
of the type
\be
\Gamma[A] = 
- \int_0^\infty \frac{dT}{T} \, e^{- m^2 T}   
\int_0^{2\pi} \frac{d\phi}{2\pi} \, e^{ i s \phi}   
\int_{_{\! PBC}}  \!\! \!\! \!\!  Dx 
\int_{_{\! TBC}}  \!\! \!\! \!\!  D{\bar c}  Dc\  \
{e}^{ -S_{gf}}
\label{gfea}
\ee
with
\be
S_{gf}
= \int_0^1 d\tau \bigg(\frac{1}{4T}  \dot x^2 +  \bar c^\alpha \dot c_\alpha  -
ig\, A^a _\mu(x)\dot x^\mu\,  \bar c^\alpha (T^a)_\alpha{}^\beta c_\beta \bigg)
\label{gfwa}
\ee
where $PBC$ denotes periodic boundary conditions, and $TBC$ twisted boundary conditions specified by
$c(1)= - e^{-i\phi}c(0)$, $\bar c(1)= - e^{i\phi}\bar c(0)$. The latter take into account the coupling to the worldline U(1)
gauge field, and arise after a field redefinition ($c(\tau)\to e^{i\phi \tau} c(\tau)$ and its complex conjugate)
that eliminates the U(1) coupling from the action while transferring it to the boundary conditions
 (for $\phi=0$ they reduce to $ABC$, antiperiodic boundary conditions, as natural for fermionic variables).

\subsection{Contribution to the $n$-gluon amplitudes}
Having found the worldline representation for the one-loop effective action due to a scalar particle, 
it is now easy to find a master formula of the Bern-Kosower type \cite{Bern:1991aq},
though containing only one-particle irreducible terms,
for the corresponding contribution to the one-loop $n$-gluon amplitudes.
This is obtained as usual  by inserting in (\ref{gfea}) a sum of plane waves
\be
A_{\mu}^{a}(x) T^a =\sum_{i=1}^n \varepsilon_\mu(k_i) e^{ik_i \cdot x}\,  T^{a_i} \;,
\ee
selecting the terms linear in each polarization  $\varepsilon_\mu(k_i)$, and choosing the generators $T^{a_i}$
in the fundamental representation. One finds 
\begin{equation}\label{n gluon scalar amplitude}
\begin{split}
\Gamma_{\rm scal}(k_1,\varepsilon_1,a_1;...;k_n,\varepsilon_n,a_n)&=-(ig)^n \int_0^\infty {dT\over T}\,e^{-m^2T} \
\int_0^{2\pi} {d\phi\over 2\pi}\,e^{is\phi}\\
&\times\int_{_{\! PBC}}  \!\! \!\! \!\!  Dx
\int_{_{\! TBC}}  \!\! \!\! \!\!  D{\bar c}  Dc\  \
{\rm e}^{ -S_2}\prod_{i=1}^n V_{\rm scal}[k_i,\varepsilon_i,a_i]
\end{split}
\end{equation}
where the gluon vertex operator is given by
\begin{equation}\label{gluon vertex scalar}
V_{\rm scal}[k,\varepsilon,a]=(T^a)_\alpha{}^\beta\int_0^1 d\tau\,\varepsilon\cdot\dot x(\tau)\,\bar c^\alpha(\tau)c_\beta(\tau)\,e^{ik\cdot x(\tau)}
\end{equation}
and $S_2$ denotes the quadratic part of the gauge fixed action \eqref{gfwa}.
On the space of periodic functions $x^\mu(\tau)$ the kinetic operator $-\frac{1}{4T}\frac{d^2}{d\tau^2}$ is not invertible, due to constant zero modes. Thus, one may split the trajectories as $x^\mu(\tau)=x^\mu_0+y^\mu(\tau)$, where
\begin{equation}
y^\mu(\tau)=\sum_{m\neq 0}y^\mu_m\,e^{2\pi im\tau}\;,\quad x^\mu_0=\int_0^1 d\tau\,x^\mu(\tau)\;,\quad \int_0^1 d\tau\,y^\mu(\tau)=0\;.
\end{equation}
The integration measure factorizes as $Dx=d^Dx_0\,Dy$, and the integration over the zero modes $x^\mu_0$
produces the delta function for momentum conservation
\be
\int d^Dx_0\,\prod_{i=1}^n\,e^{ik_i\cdot x_0}=(2\pi)^D\delta^D(k_1+k_2+...+k_n)\;.
\ee
It is also convenient to define the normalized quantum averages $\media{...}$ with respect to the quadratic action 
as $\media{f}=\frac{\int DyD\bar c Dc\,f\,e^{-S_2}}{\int DyD\bar cDc\,e^{-S_2}}$, with
\begin{equation}\label{S2 scalar}
S_2=\int_0^1 d\tau \left (\frac{1}{4T}\dot y^2+\bar c^\alpha \dot c_\alpha \right)\;,
\end{equation}
while the free path integral normalization yields
\be
\int_{_{\! PBC}}  \!\! \!\! \!\!  Dy 
\int_{_{\! TBC}}  \!\! \!\! \!\!  D{\bar c}  Dc\ e^{-S_2}
=(4\pi T)^{-D/2}\, \Det_{_{TBC}}(\de_\tau)^N=(4\pi T)^{-D/2}\left(2\cos\frac\phi2\right)^N      \;.
\ee
Also, from the action \eqref{S2 scalar} one extracts the free worldline propagators
\begin{equation}\label{2point function particle}
\media{y^\mu(\tau)y^\nu(\sigma)}=-T\,\delta^{\mu\nu}\,G(\tau-\sigma)\;,\quad\media{c_\alpha(\tau)\bar c^\beta(\sigma)}=\delta_\alpha^\beta\,\Delta(\tau-\sigma;\phi)
\end{equation}
where
\begin{equation}\label{propagators}
\begin{split}
G(\tau-\sigma) &= \left\lvert{\tau-\sigma}\right\rvert-(\tau-\sigma)^2\;,\\
\Delta(\tau-\sigma;\phi) &= \frac{1}{2\cos\frac\phi2}\Big[e^{i\frac\phi2}\theta(\tau-\sigma)-e^{-i\frac\phi2}\theta(\sigma-\tau)\Big]
\end{split}
\end{equation}
with $\theta(x)$ the step function.
A constant part in the bosonic propagator $G(\tau-\sigma)$ has been dropped, as it does not contribute in such calculations 
due to momentum conservation.  At last, one may use an exponentiation trick by writing
\be
\prod_{i=1}^n\varepsilon_i\cdot\dot y_i=\left.e^{\sum_{i=1}^n\varepsilon_i\cdot\dot y_i}\right\rvert_{{\rm lin}\,\varepsilon_1...\varepsilon_n}
\ee
with the shorthand notation $y_i=y(\tau_i)$, so that with all the previous definitions and manipulations one may
cast  the non-abelian master formula for the $n$-gluon amplitude in the following way
\begin{equation}\label{n gluon scalar master}
\begin{split}
&\Gamma_{\rm scal}(k_1,\varepsilon_1,a_1;...;k_n,\varepsilon_n,a_n) =\\
&-(ig)^n\,(2\pi)^D\delta^D(k_1+...+k_n)\,\int_0^\infty {dT\over T}\,\frac{e^{-m^2T}}{(4\pi T)^{D/2}}
\int_0^{2\pi} {d\phi\over 2\pi}\,\left(2\cos\frac\phi2\right)^{N}\,e^{is\phi}\\
&\times\int_0^1d\tau_1...\int_0^1d\tau_n\,\exp\left\{T\sum_{i,j=1}^n\,\tfrac12\,G_{ij}k_i\cdot k_j-i\,\dot G_{ij}\varepsilon_i\cdot k_j+\tfrac12\,\ddot G_{ij}\varepsilon_i\cdot\varepsilon_j\right\} \bigg \rvert_{{\rm lin}\,\varepsilon_1...\varepsilon_n}\\
&\times \left\langle  \prod_{i=1}^n(T^{a_i})_{\alpha_i}{}^{\beta_i}  \bar c^{\alpha_i}(\tau_i)c_{\beta_i}(\tau_i)\right\rangle 
\end{split}
\end{equation}
where $G_{ij}=G(\tau_i-\tau_j)$, and dots stand for derivatives with respect to the first variable. 
It is interesting to notice that the bosonic contributions yield exactly the same $\tau_i$ integrand as in the abelian case:
all the non-abelian features are captured by the extra fermionic average appearing in the last line,
and by the modular integral over $\phi$.

\subsection{The gluon self-energy}

The previous master formula can be tested by checking the scalar contribution to the gluon self-energy.
From \eqref{n gluon scalar master} one obtains 
\begin{equation}\label{se1}
\begin{split}
& \Gamma_{\rm scal}(k_1,\varepsilon_1,a_1;k_2,\varepsilon_2,a_2)=g^2\,(2\pi)^D\,\delta^D(k_1+k_2)\,{\rm tr}_{_F}(T^{a_1}T^{a_2})
\int_0^{2\pi} {d\phi\over 2\pi}\, e^{is\phi} \left(2\cos\frac\phi2\right)^{N-2} \\
& \times (\varepsilon_1\cdot\varepsilon_2\, k^2 -\varepsilon_1\cdot k\, \varepsilon_2\cdot k)\;
\int_0^\infty {dT\over T}\,\frac{e^{-m^2T}}{(4\pi T)^{D/2}} \;
T^2 \int_0^1d\tau_1\int_0^1d\tau_2\,
\dot G^2_{12}\, e^{-T k^2 G_{12}}
\end{split}
\end{equation}
where we used $\Delta(\tau-\sigma;\phi)\Delta(\sigma-\tau;\phi)=-\left(2\cos\frac\phi2\right)^{-2}$ in the fermionic contractions,
integrated by parts a term in the bosonic piece  to get a manifestly transverse amplitude, 
and used momentum conservation to define $k=k_1=-k_2$. 
It can be casted in the following form
\begin{equation}\label{self energy 2}
\Gamma_{\rm scal}(k_1,\varepsilon_1,a_1;k_2,\varepsilon_2,a_2)=(2\pi)^D\,\delta^D(k_1+k_2)\, C^{a_1a_2} \varepsilon_1^\mu\varepsilon_2^\nu \Pi_{\mu\nu}(k)
\end{equation}
to isolate the color factor 
\be 
C^{a b} =
{\rm tr}_{_F} (T^{a }T^{b} ) \int_0^{2\pi} {d\phi\over 2\pi}\, e^{is\phi} \left(2\cos\frac\phi2\right)^{N-2}
\label{color factor}
\ee
and the abelian vacuum polarization tensor  $\Pi_{\mu\nu}(k) = (\delta^{\mu\nu}k^2-k^\mu k^\nu)\, \Pi(k^2)$ 
with
\be
\Pi(k^2)= \frac{g^2}{(4\pi)^{D/2}}\,
\int_0^\infty\frac{dT}{T}e^{-m^2T}\,T^{2-\tfrac D2}
\int_0^1 du\,(1-2u)^2\,e^{-T k^2 u(1-u)} \;.
\label{vacumm pol scalar}
\ee
The latter is obtained by using the translation symmetry on the circle to fix $\tau_2=0$, so that denoting  $\tau_1=u$
one has $G(u)=u(1-u)$ and $\dot G(u)=1-2u$.  For completeness, let us proceed further by using
\be
\int_0^\infty \frac{dt}{t}\,e^{-at}\,t^z=a^{-z}\,\Gamma(z)
\ee
to perform the $T$ integral and get it into a standard form 
\begin{equation}\label{vacuum pol scalar final}
\Pi(k^2)=\frac{g^2}{(4\pi)^{D/2}}\,\Gamma\Big(2-\tfrac{D}{2}\Big)\int_0^1du\,(1-2u)^2\Big[m^2+u(1-u)k^2\Big]^{\tfrac{D}{2}-2}
\end{equation}
that is easily regulated in dimensional regularization with $D=4-2\varepsilon$ (and $\overline{\rm MS}$ scheme conventions) to get 
\be
\Pi(k^2) = \frac{1}{\varepsilon}\,\frac{g^2}{48\pi^2}-\frac{g^2}{16\pi^2}\int_0^1du\,(1-2u)^2\,\ln\frac{M^2(u,k^2)}{\mu^2}+\mathcal{O}(\varepsilon)
\ee
where $M^2(u,k^2)=m^2+k^2u(1-u)$.

In this simple example the effect of the color charge is to dress the abelian vacuum polarization by the color factor $C^{a b}$.
In the fundamental representation one has ${\rm tr}_{_F}(T^{a}T^{b}) = \frac12 \delta^{ab} $.
Using for simplicity the Wilson loop variable $z=e^{i\phi}$, and the specific value of the Chern-Simons coupling $s= 1-\frac{N}{2}$, 
one finds from (\ref{color factor})
\be
C^{ab} = \frac12 \delta^{ab} 
 \oint_{\cal C} \frac{dz}{2 \pi i} \frac{(1+z)^{N-2}}{z^{N-1}}    =  \frac12 \delta^{ab} 
 \label{cfac}
\ee
where $\cal C$ is a contour of unit radius around the origin of the $z$ complex plane.
This is the expected contribution for a particle in the fundamental representation of the color group.

It is also simple to consider other occupation numbers in the particle model to see how the color charge corresponding to the 
various antisymmetric tensor products of the fundamental representation contributes to the self energy, and in particular to 
the one-loop beta function.
One must set the Chern-Simons coupling to the value $s= n-\frac{N}{2}$, with $n=0,1,..., N$, to select the other
wave functions appearing in the expansion (\ref{wavefun}).
One finds that the integral in (\ref{cfac}) is changed to
\be 
 \oint_{\cal C} \frac{dz}{2 \pi i} \frac{(1+z)^{N-2}}{z^{N-n}} = \frac{(N-2)!}{(N-n-1)! (n-1)!}
\ee
for $1<n<N$, while it vanishes for $n=0$, as there is no residue, and for $n=N$, as there is no pole. The latter two values are
obviously correct, as the particle is colorless for such cases. 
One may verify directly the color/anticolor duality, given by changing $n\to N-n$.

Of course, if one is interested in describing particles in other representations of the color group,   
one may just insert in (\ref{1.5}) the generators in the chosen representation $R$, say of dimension $d_R$,
take the variables $c$ in that representation and $\bar c$ in its complex conjugate,
and the previous analyses goes through just by substituting  in the Chern-Simons coupling
the dimension $N$ of the fundamental representation by $d_R$.

\section{The colored spinor}

In this section we extend the previous construction to spin $1/2$ colored fermions, starting for simplicity from a phase-space approach. 
The relativistic spinning particle has an underlying $N=1$ local supersymmetry on the worldline \cite{Brink:1976sz}.
 The essential ingredient is the supercharge $Q$, that gives rise to the Dirac equation as quantum constraint,
while the hamiltonian $H$ is completely determined by the supersymmetry algebra.
For reducing to an irreducible representation of the color group we add an additional constraint $J$.
We present the massless case first, and then add a mass term by dimensional reduction.

The massless model contains as degrees of freedom the particle coordinates and momenta $x^\mu$ and $p_\mu$
together with real  Grassmann variables $\psi^\mu$ that reproduce space-time spin degrees of freedom. The
color structure of the particle is provided in exactly the same way as before by adding the $\bar c c$ sector with its coupling to the U(1)
gauge field $a$.
In minkowskian time the phase space action takes the form 
\begin{equation}\label{ph space action spinor}
S_{ph}[x,p,\psi,c,\bar c,e,\chi,a;A]=\int_0^1 d\tau
\Big[p_\mu\dot x^\mu+\tfrac{i}{2}\,\psi^\mu\dot\psi_\mu+i\bar c^\alpha\dot c_\alpha-e\,H-i\chi\,Q-a\, J \Big]
\end{equation}
where $H, Q$ and  $J$ are constraint functions that are set to vanish by the worldline gauge fields $e,\chi$ and  $a$. 
The supercharge $Q$ is defined by 
\begin{equation}\label{supercharge}
Q=\psi^\mu\,\pi_\mu=\psi^\mu\,\left(p_\mu - g\,A^a _\mu(x)\,  \bar c^\alpha\, (T^a)_\alpha{}^\beta\, c_\beta\right)
\end{equation}
where $\pi_\mu$ are gauge covariant momenta. 
It is easy to check that at the quantum level the corresponding constraint produces 
the gauge covariant massless Dirac equation
$$
 Q\ket{\Psi} =0 \quad \to \quad \Dslash\,\Psi(x) =0 \;.
$$
The hamiltonian $H$ is fixed by the supersymmetry algebra, that in terms of Poisson brackets reads
\be\label{susy H}
\{Q,Q\} = -2i\,H \quad \to \quad 
H = \tfrac12\,\pi^2+\tfrac{i}{2}\,g\,\psi^\mu\psi^\nu\,F^a_{\mu\nu}\,\bar c^\alpha\, (T^a)_\alpha{}^\beta\, c_\beta
\ee
where the non-abelian field strength is given by $F^a_{\mu\nu}=\de_\mu A_\nu^a-\de_\nu A_\mu^a + g\,f^{abc}\,A_\mu^b\,A_\nu^c $,
with $f^{abc}$ the structure constants of the gauge group.
At this stage the spinor wave function transforms in all possible tensor products of the fundamental representation, so that by
taking the U(1) current in  the form
\be
J = \bar c^\alpha c_\alpha-s\;,
\ee
that is the fermion number for the $c$ and $\bar c$ variables modified by the Chern-Simons coupling $s= n-\frac{N}{2}$, 
it implements the required projection to the sector with occupation number $n$. This completes the construction at the hamiltonian level.

We can now eliminate momenta from \eqref{ph space action spinor} by means of their equations of motion, perform a Wick rotation to 
euclidean time, and obtain the euclidean action 
\begin{equation}\label{euclidian action spinor}
\begin{split}
S
&= \int_0^1 d\tau \Big[\frac{1}{2e}\left(\dot x^\mu-\chi\psi^\mu\right)^2+\tfrac12\,\psi^\mu\dot\psi_\mu 
+\bar c^\alpha (\partial_\tau + i a)c_\alpha  - i s a\\
&-ig A^a _\mu(x)\,\dot x^\mu \, \bar c^\alpha\, (T^a)_\alpha{}^\beta\, c_\beta
+ \frac{ie}{2}\,g\,\psi^\mu\psi^\nu\,F_{\mu\nu}^a(x)\,\bar c^\alpha\, (T^a)_\alpha{}^\beta\, c_\beta \Big]\;,
\end{split}
\end{equation}
which can be used in a path integral on a circle $S^1$ to compute
the QFT effective action induced by a quark loop 
\begin{equation}
\Gamma[A] = \int_{S^1}  \
{DxD\psi D \bar c Dc De D\chi Da
\over {\rm Vol(Gauge)}}\
{e}^{- S}\;.
\end{equation}
The gauge fixing procedure goes along as in the scalar case, together with  the gauge fixing of the
local supersymmetry obtained by requiring $\chi=0$\footnote{The gravitino is antiperiodic and does not develop a modulus, 
and the associated Faddeev-Popov determinant is trivial.}.  
One may also introduce a mass term, without  breaking the supersymmetry algebra, by dimensional reduction of 
the massless theory in one dimension higher. The net effect is that, in the gauge chosen, one only finds 
an additional term $e^{-m^2T}$. Thus, the gauge fixed path integral 
reduces to
\begin{equation}\label{effective action spinor}
\Gamma[A] =
 \tfrac12\,\int_0^\infty {dT\over T}\,e^{-m^2T} \
\int_0^{2\pi} {d\phi\over 2\pi}\,e^{is\phi} \
\int_{_{\! PBC}}  \!\! \!\! \!\!  Dx
\int_{_{\! ABC}}  \!\! \!\! \!\!  D\psi
\int_{_{\! TBC}}  \!\! \!\! \!\!  D{\bar c}  Dc\  \
{e}^{ -S_{gf}}
\end{equation}
where the gauge fixed action is given by
\begin{equation}\label{GF spinor action}
\begin{split}
S_{gf} &= \int_0^1 d\tau \Big[\,\frac{1}{4T}\dot x^2+\tfrac12\,\psi^\mu\dot\psi_\mu+\bar c^\alpha \dot c_\alpha\\
&+ig\, A^a _\mu(x)\,\dot x^\mu\,  \bar c^\alpha\, (T^a)_\alpha{}^\beta\, c_\beta
-iTg\,\psi^\mu\psi^\nu\,F_{\mu\nu}^a(x)\,\bar c^\alpha\, (T^a)_\alpha{}^\beta\, c_\beta \Big]\;.
\end{split}
\end{equation}
The overall normalization of the effective action is fixed by comparing with QFT results, and includes the sign for the  loop
of a fermionic particle.

\subsection{The gluon self-energy}

In order to compute $n$-point amplitudes from \eqref{effective action spinor} we proceed in the usual way by specializing the background gauge field $A_\mu^a$ to a sum of plane waves. A complication in the spinor case is due to the non-abelian field strength 
appearing in \eqref{GF spinor action}, that yields two kind of vertices, a first one with one gluon and a second one with two gluons.
The full one-gluon vertex is given by
\begin{equation}\label{gluon vertex spinor}
V_{\rm spin}[k,\varepsilon,a]=(T^a)_\alpha{}^\beta\int_0^1 d\tau\,\Big[\varepsilon\cdot\dot x(\tau)-2iT\,k\cdot\psi(\tau)\,\varepsilon\cdot\psi(\tau)\Big]\,\bar c^\alpha(\tau)c_\beta(\tau)\,e^{ik\cdot x(\tau)}\;,
\end{equation}
and the two-gluon contact vertex by
\begin{equation}\label{2gluon contact}
V_2[k_1,\varepsilon_1,a;k_2,\varepsilon_2,b]=-2iT\,f^{abc}\,(T^c)_\alpha{}^\beta\int_0^1d\tau\,\varepsilon_1\cdot\psi(\tau)\,\varepsilon_2\cdot\psi(\tau)\,\bar c^\alpha(\tau) c_\beta(\tau)\,e^{i(k_1+k_2)\cdot x(\tau)}\;.
\end{equation}

Again, we only present the explicit calculation of the two-point function, \emph{i.e.} the spinor contribution to the gluon self-energy.
The perturbative calculation is standard, as reviewed in the previous section but with the addition of the $\psi\psi$ propagator for antiperiodic fermions that reads
\begin{equation}\label{psi propagator}
\begin{split}
&\media{\psi^\mu(\tau)\psi^\nu(\sigma)}=\delta^{\mu\nu}\,S(\tau-\sigma)\\
&S(\tau-\sigma)=\tfrac12\,\epsilon(\tau-\sigma)
\end{split}
\end{equation}
where $\epsilon(x)$ is the sign function obeying $\epsilon(0)=0$. The free path integral normalization gains an extra $2^{D/2}$ factor coming from the $\psi$ fermions, and as before we use the symbol $\media{...}$ to denote the normalized quantum average 
with respect to the quadratic action.  One can thus write the gluon self-energy as
\begin{equation}\label{self energy spinor}
\begin{split}
&\Gamma_{\rm spin}(k_1,\varepsilon_1,a_1;k_2,\varepsilon_2,a_2)=2^{D/2-1}\int_0^\infty {dT\over T}\,\frac{e^{-m^2T}}{(4\pi T)^{D/2}}
\int_0^{2\pi} {d\phi\over 2\pi}\,e^{is\phi}\left(2\cos\frac\phi2\right)^N\\
&\times\left\langle(ig)^2\prod_{i=1}^2V_{\rm spin}[k_i,\varepsilon_i,a_i]+g^2V_2[k_1,\varepsilon_1,a_1;k_2,\varepsilon_2,a_2]\right\rangle\;.
\end{split}
\end{equation}
By looking at the contact vertex \eqref{2gluon contact}, it is immediate to see that the $c\bar c$ contraction in \eqref{self energy spinor} yields ${\rm tr}_{{}_F}T^c=0$, and it is therefore irrelevant for the two-point amplitude. The computation of the other piece $\left\langle V_{\rm spin}[k_1,\varepsilon_1,a_1]V_{\rm spin}[k_2,\varepsilon_2,a_2]\right\rangle$ proceeds in the very same way as for the scalar one: the $c\bar c$ contractions give the same color factor, also one can employ the standard techniques to deal with the bosonic part, and the $\psi\psi$ Wick contractions are straightforward. Hence, as for the scalar case, we write \eqref{self energy spinor} as 
\begin{equation}\label{self energy spinor 2}
\Gamma_{\rm spin}(k_1,\varepsilon_1,a_1;k_2,\varepsilon_2,a_2)=(2\pi)^D\delta^D(k_1+k_2)\,C^{a_1a_2}\,\varepsilon^\mu_1\varepsilon^\nu_2\,\Pi_{\mu\nu}(k)\;,
\end{equation}
where the color factor $C^{ab}$ is the same as in the scalar case, eq. (\ref{color factor}), 
while the vacuum polarization 
$\Pi_{\mu\nu}(k) = (\delta^{\mu\nu}k^2-k^\mu k^\nu)\, \Pi (k^2)$ 
is now given by
\begin{equation}\label{vacuum polarization spinor}
\Pi(k^2) = -\frac{2^{D/2-1}g^2}{(4\pi)^{D/2}}\,
\int_0^\infty\frac{dT}{T}\frac{e^{-m^2T}}{(4\pi T)^{D/2}}\, T^2
\int_0^1d\tau_1\int_0^1d\tau_2\,e^{-k^2TG_{12}}\left(\dot G_{12}^2-4S_{12}^2\right)\;.
\end{equation}
One can fix $\tau_2=0$ and $\tau_1=u$, giving $S(u)=\frac12$, and performing the $T$ integral produces
\begin{equation}\label{vacuum polarization spinor final}
\Pi(k^2) = \frac{2^{D/2+1}g^2}{(4\pi)^{D/2}}\,
\Gamma\Big(2-\tfrac{D}{2}\Big)\int_0^1du\,u(1-u)\Big[m^2+u(1-u)k^2\Big]^{\tfrac{D}{2}-2}
\end{equation}
that is again easily regulated in dimensional regularization with $D=4-2\varepsilon$ to   
\be
\Pi (k^2) = \frac{1}{\varepsilon}\,\frac{g^2}{12\pi^2}-\frac{g^2}{2\pi^2}\int_0^1du\,u(1-u)\,\ln\frac{M^2(u,k^2)}{\mu^2}+\mathcal{O}(\varepsilon)
\ee
where $M^2(u,k^2)=m^2+k^2u(1-u)$.

\section*{Acknowledgments}
We wish to thank Christian Schubert for useful comments on the manuscript. 
The work of FB was supported in part by the MIUR-PRIN contract 2009-KHZKRX. 
The work of OC was partly supported by the UCMEXUS-CONACYT grant CN-12-564. 
EL acknowledges partial support of SNF Grant No. 200020-131813/1.


\begin{thebibliography}{99}

\bibitem{Bern:2008qj}
  Z.~Bern, J.~J.~M.~Carrasco and H.~Johansson,
  ``New relations for gauge-theory amplitudes,''
  Phys.\ Rev.\ D {\bf 78} (2008) 085011
  [arXiv:0805.3993 [hep-ph]].

\bibitem{Wong:1970fu}
  S.~K.~Wong,
  ``Field and particle equations for the classical Yang-Mills field and particles with isotopic spin,''
  Nuovo Cim.\ A {\bf 65} (1970) 689.

\bibitem{Balachandran:1976ya}
  A.~P.~Balachandran, P.~Salomonson, B.~-S.~Skagerstam and J.~-O.~Winnberg,
  ``Classical description of particle interacting with nonabelian gauge field,''
  Phys.\ Rev.\ D {\bf 15} (1977) 2308.

\bibitem{Barducci:1976xq}
  A.~Barducci, R.~Casalbuoni and L.~Lusanna,
  ``Classical scalar and spinning particles interacting with external Yang-Mills fields,''
  Nucl.\ Phys.\ B {\bf 124} (1977) 93.

\bibitem{Berezin:1976eg}
  F.~A.~Berezin and M.~S.~Marinov,
  ``Particle spin dynamics as the Grassmann variant of classical mechanics,''
  Annals Phys.\  {\bf 104} (1977) 336.

\bibitem{Brink:1976sz} 
  L.~Brink, S.~Deser, B.~Zumino, P.~Di Vecchia and P.~S.~Howe,
  ``Local supersymmetry for spinning particles,''
  Phys.\ Lett.\ B {\bf 64}, 435 (1976).
 
\bibitem{Feynman:1951gn}
  R.~P.~Feynman,
  ``An operator calculus having applications in quantum electrodynamics,''
  Phys.\ Rev.\  {\bf 84} (1951) 108.

\bibitem{Strassler:1992zr}
  M.~J.~Strassler,
  ``Field theory without Feynman diagrams: One loop effective actions,''
  Nucl.\ Phys.\ B {\bf 385} (1992) 145
  [hep-ph/9205205].

\bibitem{Bastianelli:1992ct}
  F.~Bastianelli and P.~van Nieuwenhuizen,
  ``Trace anomalies from quantum mechanics,''
  Nucl.\ Phys.\ B {\bf 389} (1993) 53
  [hep-th/9208059].
  
\bibitem{Reuter:1996zm}
  M.~Reuter, M.~G.~Schmidt and C.~Schubert,
  ``Constant external fields in gauge theory and the spin 0, 1/2, 1 path integrals,''
  Annals Phys.\  {\bf 259} (1997) 313
  [hep-th/9610191].

\bibitem{Sato:1998sf}
  H.~-T.~Sato and M.~G.~Schmidt,
  ``Worldline approach to the Bern-Kosower formalism in two loop Yang-Mills theory,''
  Nucl.\ Phys.\ B {\bf 560} (1999) 551
  [hep-th/9812229].

\bibitem{Schubert:2001he}
  C.~Schubert,
  ``Perturbative quantum field theory in the string inspired formalism,''
  Phys.\ Rept.\  {\bf 355} (2001) 73
  [hep-th/0101036].
  
\bibitem{Dai:2008bh}
  P.~Dai, Y.~-t.~Huang and W.~Siegel,
  ``Worldgraph approach to Yang-Mills amplitudes from N=2 spinning particle,''
  JHEP {\bf 0810} (2008) 027
  [arXiv:0807.0391 [hep-th]].
  
\bibitem{Ahmadiniaz:2012xp}
  N.~Ahmadiniaz and C.~Schubert,
  ``A covariant representation of the Ball-Chiu vertex,''
  Nucl.\ Phys.\ B {\bf 869} (2013) 417
  [arXiv:1210.2331 [hep-ph]].
  
\bibitem{Ahmadiniaz:2012ie}
  N.~Ahmadiniaz, C.~Schubert and V.~M.~Villanueva,
  ``String-inspired representations of photon/gluon amplitudes,''
  JHEP {\bf 1301} (2013) 132
  [arXiv:1211.1821 [hep-th]].
  
\bibitem{D'Hoker:1995bj}
  E.~D'Hoker and D.~G.~Gagne,
  ``Worldline path integrals for fermions with general couplings,''
  Nucl.\ Phys.\ B {\bf 467} (1996) 297
  [hep-th/9512080].

\bibitem{JalilianMarian:1999xt}
  J.~Jalilian-Marian, S.~Jeon, R.~Venugopalan and J.~Wirstam,
  ``Minding one's P's and Q's: From the one loop effective action in quantum field theory to classical transport theory,''
  Phys.\ Rev.\ D {\bf 62} (2000) 045020
  [hep-ph/9910299].

\bibitem{Khorsand:2001dx}
  P.~Khorsand and T.~R.~Taylor,
  ``Renormalization of boundary fermions and world volume potentials on D-branes,''
  Nucl.\ Phys.\ B {\bf 611} (2001) 239
  [hep-th/0106244].

\bibitem{Howe:1989vn}
  P.~S.~Howe, S.~Penati, M.~Pernici and P.~K.~Townsend,
  ``A particle mechanics description of antisymmetric tensor fields,''
  Class.\ Quant.\ Grav.\  {\bf 6} (1989) 1125.

\bibitem{Bastianelli:2005vk}
  F.~Bastianelli, P.~Benincasa and S.~Giombi,
  ``Worldline approach to vector and antisymmetric tensor fields,''
  JHEP {\bf 0504} (2005) 010
  [hep-th/0503155].
    
\bibitem{Bastianelli:2011pe} 
  F.~Bastianelli and R.~Bonezzi,
  ``Quantum theory of massless (p,0)-forms,''
  JHEP {\bf 1109}, 018 (2011)
  [arXiv:1107.3661 [hep-th]].
    
\bibitem{Bastianelli:2009eh}
  F.~Bastianelli, O.~Corradini and A.~Waldron,
  ``Detours and paths: BRST complexes and worldline formalism,''
  JHEP {\bf 0905} (2009) 017
  [arXiv:0902.0530 [hep-th]].
  
\bibitem{Bastianelli:2012nh} 
  F.~Bastianelli, R.~Bonezzi and C.~Iazeolla,
  ``Quantum theories of (p,q)-forms,''
  JHEP {\bf 1208}, 045 (2012)
  [arXiv:1204.5954 [hep-th]].
  
\bibitem{Bastianelli:2013tsa}
  F.~Bastianelli and R.~Bonezzi,
  ``One-loop quantum gravity from a worldline viewpoint,''
  JHEP {\bf 1307} (2013) 016
  [arXiv:1304.7135 [hep-th]].

\bibitem{Bern:1991aq}
  Z.~Bern and D.~A.~Kosower,
  ``The computation of loop amplitudes in gauge theories,''
  Nucl.\ Phys.\ B {\bf 379} (1992) 451.
  
\end{thebibliography}
\end{document}